\documentclass[twocolumn]{aastex62}

\usepackage{natbib}
\usepackage{hyperref}
\usepackage{floatrow}
\usepackage{amsmath,amssymb}
\usepackage{epstopdf}
\usepackage{ulem}
\accepted{\today}

\submitjournal{ApJ}

\shorttitle{Wolf-Rayet Bubble G2.4+1.4}
\shortauthors{Prajapati et al.}

\begin{document}

\title{Investigating Particle Acceleration in the Wolf-Rayet Bubble G2.4+1.4}

\correspondingauthor{Anandmayee Tej}
\email{tej@iist.ac.in}

\author{Prachi Prajapati}
\affil{Indian Institute of Space Science and Technology, Trivandrum, India}

\author{Anandmayee Tej}
\affiliation{Indian Institute of Space Science and Technology, Trivandrum, India}

\author{Santiago del Palacio}
\affiliation{Instituto Argentino de Radioastronom\'ia, Argentina}

\author{Paula Benaglia}
\affiliation{Instituto Argentino de Radioastronom\'ia, Argentina}

\author{Ishwara-Chandra CH}
\affiliation{National Centre for Radio Astrophysics, Pune, India}

\author{Sarita Vig}
\affiliation{Indian Institute of Space Science and Technology, Trivandrum, India}

\author{Samir Mandal}
\affiliation{Indian Institute of Space Science and Technology, Trivandrum, India}

\author{Swarna Kanti Ghosh}
\affiliation{Tata Institute of Fundamental Research, Mumbai, India}

%============================================================

\begin{abstract}
% no more than 250 words! (180 so far)
The supersonic winds produced by massive stars carry a large amount of kinetic power. In numerous scenarios such winds have been proven to produce shocks in which relativistic particles are accelerated emitting non-thermal radiation. Here, we report the first detection of non-thermal emission from a single stellar bubble, G2.4+1.4, associated with a WO star. We observed this source with the uGMRT in Band 4 ($550-850$~MHz) and Band 5 ($1050-1450$~MHz). We present intensity and spectral index maps for this source that are consistent with synchrotron emission (average spectral index, $\alpha = -0.83 \pm 0.10$). The fraction of the available kinetic wind power that is converted into cosmic ray acceleration is estimated to be of the order of a few per cent. This finding constitutes an observational breakthrough and gives new insight on the non-thermal physical processes taking place in the environments of isolated massive stars. 
In particular, our results show that non-runaway isolated massive stars are capable of accelerating relativistic particles and are therefore confirmed as sources of Galactic cosmic rays. 
%In particular, our results confirm that non-runaway isolated massive stars are cosmic ray injectors. 
\end{abstract}

%============================================================
%Main Text – no more than 3500 words (not including appendices or other supplementary material)
%============================================================
%
\section{Introduction} \label{sec:intro}
%
%============================================================

Massive stars profoundly influence the surrounding interstellar medium (ISM) through their mechanical, radiative, and chemical feedback. In recent years, stellar systems involving massive OB and Wolf-Rayet (WR) stars have unraveled interesting laboratories to investigate particle acceleration in strong shocks related to the powerful stellar winds. The detection of non-thermal (NT) emission in such systems imply that relativistic particles can be accelerated in them, most likely via diffusive shock acceleration \citep[DSA;][and references therein]{Drury1983}. This acceleration mechanism is ubiquitous in sources with supersonic outflows, such as supernovae remnants \citep[e.g.][]{2015A&ARv..23....3D}, Herbig-Haro objects \citep{2019MNRAS.482.4687R}, proto-stellar jets \citep{2018MNRAS.474.3808V}, compact young stellar clusters \citep{2018A&A...611A..77Y}, and microquasars \citep{Mirabel1994}. In OB and WR systems, the detection of synchrotron emission is strongly correlated with binarity \citep{2000MNRAS.319.1005D, 2006MNRAS.371.1280D}. In these massive binaries, standing shocks result from the collision of the stellar winds, which can lead to efficient particle acceleration \citep[e.g.][]{Eichler1993,DeBecker2007}. Numbered at 40 odd, these are christened as particle-accelerating colliding-wind binaries \citep{2013A&A...558A..28D}. In addition, stellar bow shocks produced by runaway massive stars could also be efficient at accelerating particles \citep[e.g.,][]{delValle2012, delValle2014, delPalacio2018}, although the only evidence so far is the detection of NT emission in the system BD+43$^\circ$3654 \citep{2010A&A...517L..10B}.
In consequence, massive stars are now suggested as major factories of Galactic cosmic rays \citep[CRs;][]{2018JKAS...51...37S}, perhaps even the dominant accelerators of CRs up to PeV energies \citep{2018A&A...611A..77Y}. 

A missing link in the assessment of massive stars as progenitors of Galactic CRs is the role of isolated stars, in particular those that do not belong to the runaway class. The powerful stellar winds generate strong shocks in the surrounding material as they propagate, piling it up and sculpting the environment into a stellar bubble. %Occurrence of 
Both termination shocks and shocks in co-rotating interaction regions at the base of the wind are conducive for relativistic particle acceleration. However, until date no study has reported the detection of NT emission in such isolated objects. 
Indirect evidence of the presence of strong shocks can be gathered in X-rays, given that such shocks heat the plasma to very high temperatures. However, in stellar bubbles the turbulent mixing and thermal conduction produce gas of intermediate temperature and density, which only emits soft and diffuse X-rays as has been observed in four WR nebulae \citep{Toala2018}.

In this paper we investigate the radio emission from the stellar bubble around WR~102. Located at a distance of 2.88~kpc \citep{2018arXiv180704293S}, WR~102 is a single, oxygen-rich WR star of the rare spectral type WO2. These stars represent a very short evolutionary phase of very massive stars (40--60~M$_\odot$) and are the likely progenitors of Type Ic supernova \citep{2015IAUS..307..144T}. Remarkably, these very peculiar objects (there are only four detected WO stars in our Galaxy, including WR~102) have the fastest winds with $v_\mathrm{w} = 4500-7000$~km~s$^{-1}$ \citep{1992A&A...265..563P}. This fact makes WO stars excellent candidates for the search of NT emission as the wind velocity has a double impact on the NT particle distribution. On the one hand, the energy budget for NT particle acceleration depends on the wind kinetic power, which for a star with mass-loss rate $\dot{M}$ is $P_\mathrm{kin} \approx 0.5~\dot{M}~v_\mathrm{w}^2$. On the other hand, under the assumption of Bohm diffusion, the rate at which relativistic particles are accelerated scales with the square of the shock (and therefore the wind) velocity \citep[e.g.][]{Drury1983}. Thus, the filamentary nebula G2.4+1.4 around WR~102, with $v_\mathrm{w} = 5000$~km~s$^{-1}$ and $\dot{M} = 5.4 \times 10^{-6} \rm M_{\odot}$~yr$^{-1}$ \citep{2018arXiv180704293S}, is an ideal target to search for NT radio emission.

The hydrodynamics of stellar bubbles are complex as the wind properties vary throughout the stellar evolution from main-sequence to the WR phase. This is evident in the complex and striking morphology seen in G2.4+1.4 in the optical, infrared and radio \citep{1982ApJ...254..132T, 1987MNRAS.225..221G, 1990ApJ...351..563D, 1995ApJ...439..637G, 2015A&A...578A..66T}. The `looped' or `multi-ringed' filamentary morphology revealed in all wavelengths points to possible interaction/collision regions. 
%=============================================================
%
\section{Radio Observations} \label{sec:obs}
%
%=============================================================

Low-frequency radio continuum mapping of G2.4+1.4 was carried out with the upgraded Giant Meterwave Radio Telescope (uGMRT), India. GMRT offers a hybrid, Y-shaped configuration consisting of thirty antennas. The three arms contain six antennas each and twelve antennas are placed randomly within a central, compact region of one square kilometre. The largest baseline probed with GMRT is $\sim$ 25~km and the shortest spacing is $\sim$ 100~m, providing a configuration that allows to study structures at various spatial scales. An overview of GMRT systems can be found in \cite{1991CuSc...60...95S} and the details of the upgradation are presented in \cite{article}. G2.4+1.4 was observed in Band~4 ($550-850$~MHz) and Band~5 ($1050-1450$~MHz) which have better sensitivity and {\it uv} coverage compared to the {\it Legacy} narrow-band set up. Primary calibrators 3C286 and 3C48 were observed for flux and bandpass calibration. To correct for the phase and amplitude variation over the duration of the observations, phase calibrators 1822-096 (Band 4) and 1751-253 (Band 5) were observed for 5~min after each 40~min scan of the target.
Standard procedures of data reduction using \textit{Astronomical Image Processing System} (AIPS) were implemented to generate the continuum maps. Calibrated data in each band were split into sub-bands of bandwidth $\sim$ 32~MHz and carefully scrutinized to identify and flag out bad spectral channels (due to RFI) and data points (due to non-functional antennas). We retain five clean sub-bands each in Band 4 and Band 5, with central frequencies of 605, 640, 675, 710, 745~MHz and 1297, 1327, 1361, 1395, 1429~MHz, respectively. Each sub-band was imaged separately employing channel averaging and wide-field imaging technique to account for bandwidth smearing and {\it w-term} effects, respectively. While imaging, the beam in all sub-bands was matched to yield a resolution of $15\arcsec \times 15\arcsec$, which helps in smoothing the maps and removing small scale statistical fluctuations, if present. Several iterations of self-calibration were carried out to correct for the phases and improve the image quality. Primary beam correction was applied to generate the final images.
  
G2.4+1.4 is located near the Galactic plane, whereas the flux calibrators observed are at higher Galactic latitudes. Therefore, a correction to the system temperature was required to account for the contribution from Galactic diffuse emission, which may become significant at the low frequencies of our observations. Since measurements of the variations in the system temperatures of the antennas at the GMRT are not automatically implemented during observations, correction factors were determined using the 408~MHz all-sky continuum survey of \citep{1982A&AS...47....1H} and assuming that the Galactic diffuse emission follows a power-law spectrum with a slope of $-2.55$ \citep{1999A&AS..137....7R}. The correction factors lie between 1.17 and 2.08 (highest value occurring for the lowest frequency) and are used to scale the final images.

%=================================================================================
%
\section{Nature of the radio emission} \label{sec:ionized_emission}
%
%=================================================================================

The primary focus of this study is to ascertain the nature of the radio emission in G2.4+1.4. For this, we derive the spectral index which is a key parameter that enables to decipher the underlying physics of the source and the radiation mechanism(s).

%----------------------------------------
\subsection{Morphology of the source}
\begin{figure*}
\centering
%\plotone{./uGMRT_maps.eps}
\includegraphics[width=0.89\textwidth]{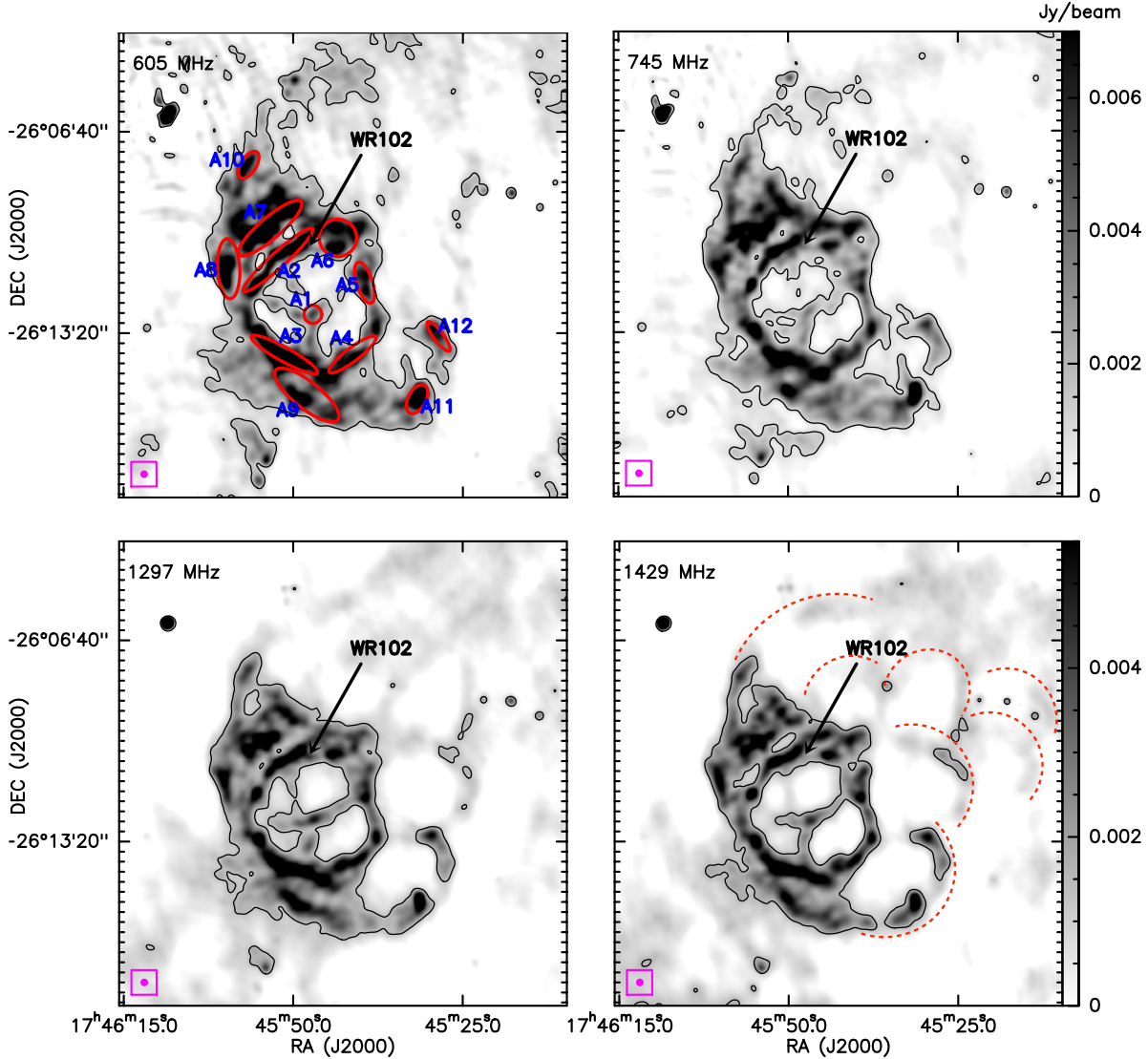}
\caption{uGMRT radio continuum maps of G2.4+1.4 at 605 and 745~MHz from Band 4, and 1297 and 1429~MHz from Band 5, with 3$\sigma$ contours overlaid ($rms$ values for each sub-band are given in Table~\ref{tab:flux-density}). The beam ($15\arcsec \times 15\arcsec$) is shown as circles at bottom left corner of the maps. Some of the faint  discernible loops are traced with dotted red curves in the 1429~MHz map. The arrow points to the position of WR~102. Identified elliptical apertures are shown in the 605~MHz map.}
\label{fig:radio_maps}
\end{figure*}

%---------------------------------------
The uGMRT radio continuum maps generated in all ten sub-bands display the characteristic multi-ringed, filamentary morphology previously mentioned. Figure~\ref{fig:radio_maps} shows two sub-bands from each Band 4 and Band 5. The maps display a central region that appears to be connected to a spherical shell via three spoke-like filaments. Beyond this, a network of loops forming an outer shell and external rims are visible. The nebula is extended in the north-west direction where faint loops are discernible in the maps and indicated in Figure~\ref{fig:radio_maps} (1429~MHz map) with dotted red curves. Noticeable in the maps is the striking asymmetry seen in the bubble structure and the location of WR 102 that is offset from the centre.

The interesting morphology of G2.4+1.4 has been debated in previous studies. \cite{1990ApJ...359..419D} attributed the striking morphology to the presence of large scale Rayleigh-Taylor (RT) instabilities. Their hypothesis places WR~102 at the edge of a molecular cloud that stalls the expansion on one side (south-east) whereas the stellar wind inflates the bubble in the other (north-west) direction into the low-density ISM. RT instabilities arise as the expansion happens along a strong density gradient. This results in the characteristic `scalloping' morphology. However, the presence of the proposed molecular cloud was not confirmed from CO observations carried out by \cite{2008RMxAC..33..140A}. An alternate scenario was presented by \cite{1995IAUS..163...70B, 1995MNRAS.273..443B}. These authors carried out detailed simulations and advocate for a moving central star that would adequately explain the observed morphology. 

%

%----------------------------------------
\subsection{Spectral index}
%---------------------------------------

The spectral index $\alpha$ is defined as $S_\nu \propto \nu^{\alpha}$, where $S_\nu$ is the flux density at frequency $\nu$. Ionized gas emits thermal {\it free-free} emission with spectral index ranging from $+2$ (optically thick) to $-0.1$ (optically thin). Given that the opacity is frequency-dependent, a source can switch from being optically thick at low frequencies to optically thin at high frequencies. Moreover, the opacity is strongly dependent on the density of the medium, hence the turnover frequency can differ for different sources. In comparison, relativistic electrons emit synchrotron radiation in the radio domain with a typical intrinsic spectral index $\alpha \sim -0.5$, although it could be less negative or even positive due to absorption/suppression effects \citep{Melrose1980}. In any case, a value of $\alpha < -0.1$ is conclusive evidence for NT emission, whereas positive values of the spectral index usually suggest a thermal source. 
In astrophysical sources it is most likely that a thermal plasma coexists with the relativistic population of electrons (if present), and therefore the observed spectra would have contributions from both the thermal and NT components. Nonetheless, it is possible to separate the two components by sampling the SED over a large range of frequencies. In particular, the NT emission is expected to dominate the spectra at low frequencies given that it has a negative spectral index, whereas at higher frequencies thermal radiation 
prevails \citep[see e.g.][]{DeBecker2018}.
In this regard, the high-sensitivity new uGMRT data presented in this paper probes the ideal low-frequency domain appropriate to reveal unambiguous evidence of NT radiation. 

%----------------------------------------
\subsection{Observed spectral index maps}
%----------------------------------------

For a reliable estimate of the spectral index it is critical to ensure that same spatial scales are probed at the frequencies sampled. Given that GMRT is not a scaled array, we generate the maps in the {\it uv} range (0.164 -- 49.5~k$\lambda$) common to all sub-bands. Flux density integrated within the $3\sigma$ contour along with the {\it rms} noise in each sub-band map are listed in Table~\ref{tab:flux-density}. Uncertainties in the estimated flux densities are calculated using the expression from \cite{Sanchez-Monge2013}, 
%$\rm \left[ {\left( 2\sigma \sqrt{{\theta}_{src}/{\theta}_{bm}} \right)}^{2}+{(2 {\sigma}^{\prime})^2 \right]^{1/2}}$
$\rm \sqrt{{\left( 2\sigma \sqrt{{\theta}_{src}/{\theta}_{bm}} \right)}^2+{(2 {\sigma}^{\prime})}^2}$, where $\sigma$ is the {\it rms} noise level of the map, ${\sigma}^{\prime}$ is the error in flux scale calibration, $\theta_\mathrm{bm}$ represents the size of the beam, and $\theta_\mathrm{src}$ is the source size. The uncertainty in the flux calibration of GMRT is taken to be $5\%$ \citep{10.1111/j.1365-2966.2006.11225.x}. The measured flux density values yield a global spectral index of  $-0.83 \pm 0.10$. A similar power law slope is also seen within the two bands. 

%%%%%%%%%%% beginning deluxetable %%%%%%%%%%%%%%%%%%%%%%
\begin{deluxetable}{ccccc}
\tablecaption{Integrated flux density of G2.4+1.4 and {\it rms} noise in uGMRT sub-bands. Flux density values listed are integrated within $3\sigma$ contour in each sub-band. Refer to the text for the error estimation expression used.
\label{tab:flux-density}}
\tablehead{\colhead{Frequency } & \colhead{Integrated flux density} & \colhead{\textit{rms} noise} \\ 
\colhead{(MHz)} & \colhead{(Jy)} & \colhead{(mJy/beam)} }
\startdata
605 & 2.60$\pm$0.26 & 0.46 \\
640 & 2.54$\pm$0.25 & 0.51 \\
675 & 2.23$\pm$0.22 & 0.34 \\
710 & 2.17$\pm$0.22 & 0.43 \\
745 & 1.99$\pm$0.20 & 0.34 \\
1297 & 1.34$\pm$0.14 & 0.44 \\
1327 & 1.31$\pm$0.13 & 0.47 \\
1361 & 1.30$\pm$0.13 & 0.36 \\
1395 & 1.26$\pm$0.13 & 0.38 \\
1429 & 1.21$\pm$0.12 & 0.49 \\
\enddata
\end{deluxetable}
%%%%%%%%%%% ending deluxetable %%%%%%%%%%%%%%%%%%%%%%

\begin{figure*}
    \plotone{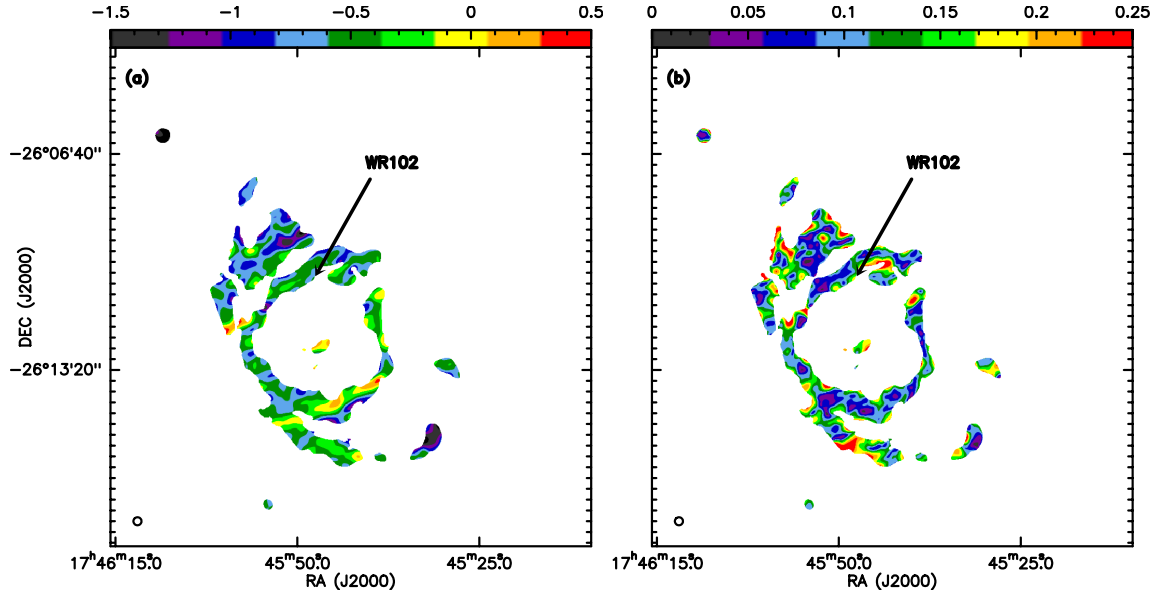}
    \caption{(a) uGMRT  Band~4 $-$ Band~5 spectral index map of G2.4+1.4 constructed for pixels above $5\sigma$ level in all sub-bands. (b) The corresponding error map. The beam ($15\arcsec \times 15\arcsec$) is shown as a circle at bottom left corner. The arrow points to the position of WR102.}
    \label{fig:SI_map}
\end{figure*}

To probe the variation of the spectral index across the bubble, we identify apertures and estimate the spectral index from the integrated flux density within these. Table~\ref{tab:aperture_SI} gives the estimated values which lie between $-1.19$ and $-0.18$. We also construct a spectral index map of G2.4+1.4, shown in Figure~\ref{fig:SI_map}, by retaining only those pixels with flux density values greater than $5\sigma$ ($\sigma$ being the {\it rms} noise of the map) in all the ten sub-band maps. The range of spectral index values seen in this map lie between $-1.7$ and $0.5$ with errors typically within 0.2.

%%%%%%%%%%% beginning deluxetable %%%%%%%%%%%%%%%%%%%%%%
\begin{deluxetable}{cc}
\tablecaption{ Spectral index estimates within identified apertures of G2.4+1.4.
\label{tab:aperture_SI}}
\tablehead{\colhead{Aperture } & \colhead{Spectral Index ($\alpha$)} } 
\startdata
Central region: &  \\ 
A1 &  $-0.18\pm0.14$\\ 
\tableline
Inner shell: &  \\ 
A2 &   $-0.51\pm0.04$ \\ 
A3 &   $-0.51\pm0.05$ \\ 
A4 &   $-0.30\pm0.04$ \\ 
A5 &   $-0.42\pm0.06$ \\ 
A6 &   $-0.57\pm0.05$ \\ 
\tableline
Outer shell: &  \\
A7  &   $-0.71\pm0.02$ \\
A8  &   $-0.56\pm0.04$ \\ 
A9  &   $-0.46\pm0.05$ \\ 
\tableline
External rims: &  \\ 
A10 &   $-0.78\pm0.05$ \\ 
A11 &   $-1.19\pm0.05$ \\ 
A12 &   $-0.55\pm0.09$ \\
\enddata
\end{deluxetable}
%%%%%%%%%%% ending deluxetable %%%%%%%%%%%%%%%%%%%%%%

%%%%%%%%%%% beginning figure %%%%%%%%%%%%%%%%%%%%%%

Thermal radio emission from a single stellar wind is produced by the {\it free-free} mechanism and it exhibits a spectral index $\alpha \sim 0.6$ \citep{1975MNRAS.170...41W}. In G2.4+1.4, the global and spatially resolved spectral index estimations are significantly lower than $-0.1$, which is a compelling evidence of NT emission. This enigmatic nebula has been a target of interest for over five decades. The nature of its radio emission has been debated by several authors. \cite{1968ApJ...154L..75G} were the first to report the 20~cm (1.4~GHz) and 11~cm (2.7~GHz) radio observations of G2.4+1.4. These authors assumed a thermal source but did not rule out the possibility of NT emission. A more detailed study was presented by \cite{1975ApJ...198..111J}, who investigated the nature of this source based on high-frequency (15.5 and 31.4~GHz) observations and flux density measurements compiled from literature. The spectrum was found to be clearly NT, and hence G2.4+1.4 was proposed to be a supernova remnant. A quantitative estimate based on the values listed in Table~1 of \cite{1975ApJ...198..111J} yields a spectral index $\alpha = -0.7$. This conclusion was supported by \cite{1982ApJ...254..132T} based on their optical narrow-band imaging and Fabry-Perot spectroscopic observations. With additional observations at 4.86~GHz using the VLA, \cite{1987MNRAS.225..221G} re-visited the debate and extensively discussed the possible uncertainties associated with the flux densities used by \cite{1975ApJ...198..111J}. These authors did not rule out NT emission but favoured a thermal wind-blown bubble instead. G2.4+1.4 was observed as part of the Effelsberg radio continuum surveys at 1408 and 2695~MHz \citep{1984A&AS...58..197R, 1990A&AS...85..633R}. The flux density measurements from these observations also yielded a NT spectral index of $\alpha = -0.7$. However, in a subsequent study, \citet{1994MNRAS.270..835G} used MOST-843~MHz and the Effelsberg-2695~MHz and calculated $\alpha = -0.13$, which they interpreted as optically thin free-free emission, thus negating a NT origin. Similar results were presented 
by \cite{1995ApJ...439..637G}, using MOST-843~MHz and VLA-1.49~GHz data, who propounded a thermal mass-loss bubble picture for G2.4+1.4. However, given that the claim of the thermal nature of the source was based on flux density measurements from Effelsberg, VLA, and MOST maps, different sensitivities and {\it uv} coverage would render the estimated spectral index values uncertain. The spectral index estimate from the Effelsberg data ($\alpha = -0.7$) is in close agreement with the uGMRT results ($\alpha \sim -0.8$) presented here, for which the flux density measurements are not affected by the above mentioned uncertainties and therefore provide high-confidence and conclusive evidence of NT emission.   

From the spectral index values of the identified apertures listed in Table~\ref{tab:aperture_SI}, we see a hint of softening of the spectra (i.e more negative spectral indices) as we move from the centre to the external rims of the bubble. 
A few scenarios could be invoked to explain this trend. One possibility could be the contamination from thermal emission in the central, denser region where the {\it free-free} emission is higher. The other picture could involve the diffusion of particles which makes the particle energy distribution softer and therefore the spectral index values more negative. Further studies involving detailed modelling are necessary before one can conclusively interpret the observed trend. 

%=================================================================================
%
\section{Non-thermal radiative model} \label{sec:energetics}
%
%=================================================================================

The value of the magnetic field strength in the emitter and its energetics are encoded in the observed SED. Despite this, it is not possible to disentangle such unknowns if only radio data is available. However, we can infer (or at least constrain) them via rather simple and reliable modelling. We develop a one-zone model to estimate the magnetic field of %the synchrotron source for 
the stellar bubble associated with G2.4+1.4. The model is based on the approach discussed in \cite{delPalacio2018} for stellar bow shocks and is globally consistent with the standard procedure for estimating magnetic fields in NT sources \cite[e.g.][]{1979rpa..book.....R, 1992hea..book.....L, DeBecker2018}. Here, the detailed structure of the synchrotron emitter is neglected and it is considered to be homogeneous. Since G2.4+1.4 has a nearly spherical morphology, we assume a spherical synchrotron emitting volume of radius $\sim 3\arcmin$ ($\sim 2.51$~pc at a distance of 2.88~kpc). We note, nonetheless, that the exact shape and volume of the region do not have a strong impact in the results \cite[e.g.][]{DeBecker2018}.

Charged relativistic particles moving in the presence of a magnetic field emit synchrotron radiation. This process is much more efficient for electrons than for protons, so hereafter we refer only to electron synchrotron radiation. The same observed synchrotron luminosity can be achieved either by a less energetic electron population in a strong magnetic field or by more energetic electrons in a weak magnetic field. To account for this degeneracy between the amount of energy injected into NT particles and the intensity of the magnetic field, we introduce two parameters in the model related to how the energy is distributed between the magnetic field and the different species of relativistic particles\footnote{Alternatively, one can define $\eta_\mathrm{e}= \frac{U_\mathrm{e}}{U_\mathrm{NT}}$, which is equivalent to  $\eta_\mathrm{e} = \frac{K_\mathrm{ep}}{1+K_\mathrm{ep}}$ \citep[e.g.][]{DeBecker2018}.}:
\begin{align}
    \eta_\mathrm{mag} &= \frac{U_\mathrm{NT}}{U_\mathrm{mag}},\\
    K_\mathrm{ep} &= \frac{U_\mathrm{e}}{U_\mathrm{p}},
\end{align}
where $U_\mathrm{mag}=B^2/(8\pi)$, $U_\mathrm{e}$ and $U_\mathrm{p}$ are the energy densities in the magnetic field, NT electrons and NT protons, respectively, and $U_\mathrm{NT} = U_\mathrm{e} + U_\mathrm{p}$. We adopt a value of $K_\mathrm{ep}=0.01$, consistent with DSA \citep[see][and references therein for a discussion on electron-to-ion luminosity ratios]{Merten:2017mzg}. Considering that the relativistic electron energy distribution is given by $N_\mathrm{e}(E)$, we have:
\begin{equation}
    U_\mathrm{e} = \frac{1}{V} \int_{E_{min}}^{E_\mathrm{max}} E N_\mathrm{e}(E) \mathrm{d}E \; ,
\end{equation}
where $V$ is the emitter volume. In order to calculate the particle energy distribution, $N_\mathrm{e}(E)$, in a self-consistent fashion, we solve the transport equation for a stationary and homogeneous emitter. 

%To compute $N_\mathrm{e}(E), $
First, we have to characterize the injected particle distribution, $Q(E)$. Different particle acceleration mechanisms, including DSA, inject relativistic particles with a power-law distribution, $Q(E) = Q_0 E^{-p} \exp{(-E/E_\mathrm{c})}$, where $Q_0$ is a normalization constant, $p$ is the distribution spectral index, and $E_\mathrm{c}$ is the cutoff energy. The value of $p$ can be derived from the radio spectral index ($\alpha = -0.83 \pm 0.1$) as $p=-2\alpha+1 \simeq 2.66$. The value of $E_\mathrm{c}$ is given by the balance between energy gain and loss processes, and it is obtained by equating the characteristic acceleration and synchrotron times, as the latter completely dominates the energy losses for high-energy particles in this scenario. The acceleration time is $t_\mathrm{acc} = \eta_\mathrm{acc}\,E/(B\,c\,q)$~s, with $q$ the elementary charge and $\eta_\mathrm{acc}$ the acceleration efficiency, taken as $\eta_\mathrm{acc} = 2 \pi (c/v_\mathrm{w})^2$ in the Bohm diffusion regime \citep[though different values of this parameter related to uncertainties in the actual diffusion coefficient are also possible,][]{Drury1983}.

The normalization constant $Q_0$ is obtained from the condition $P_\mathrm{NT} = \int_{E_\mathrm{min}}^{E_\mathrm{max}} E \, Q(E) \, \mathrm{d}E$, where $E_\mathrm{min}$ and $E_\mathrm{max}$ are set to 1~MeV and $10 \, E_\mathrm{c}$, respectively, and $P_\mathrm{NT}$ is the power injected in relativistic particles. The latter is taken as a fraction $\chi_\mathrm{NT}$ of the total kinetic power in the stellar wind,
\begin{equation}
    P_\mathrm{w} = 3.2 \times {10}^{35} \left(\frac{\dot{M}}{M_{\odot} \, \mathrm{yr}^{-1}} \right) \cdot {\left(\frac{v_\infty}{\mathrm{km}~\mathrm{s}^{-1}} \right)}^{2} \sim 4.2 \times {10}^{37} \; \mathrm{erg~s}^{-1} \;,
\end{equation}
such that $P_\mathrm{NT} = \chi_\mathrm{NT} \, P_\mathrm{w}$. The value of $\chi_\mathrm{NT}$ gives an idea of how efficiently energy is converted to NT particles. For example, typical values of $\chi_\mathrm{NT} \simeq 0.1$ are found in the literature for supernova remnants \citep{1985ICRC....3..136D, 2011Ap&SS.336..257R}.

In a stationary scenario, as it is the case with G2.4+1.4 (at least for timescales of years), the solution to the transport equation for an injected electron distribution $Q(E)$ under radiative cooling (dominated by synchrotron losses) and non-radiative cooling (mechanical work exerted to the surrounding medium, often referred to as adiabatic losses) can be calculated as 
\begin{equation}
    N(E)=\frac{1}{\mid{\dot{E}}\mid} \int_{E}^{E_\mathrm{max}} Q(E^{\prime}) \mathrm{d}E^{\prime} \; ,
\end{equation}
 where $\mid{\dot{E}}\mid=E/t_\mathrm{cool}$ represents the total (radiative plus non-radiative) energy losses. The synchrotron cooling time, $t_\mathrm{sy}(E,B) \propto E^{-1} B^{-2}$, can be calculated precisely, whereas the characteristic adiabatic loss time is approximately $t_\mathrm{adi} \sim R/v_\mathrm{w}$, with $R \approx 2.5$~pc the characteristic size of the bubble.

\begin{figure}
    \centering
    \includegraphics[angle=270, width=0.8\linewidth]{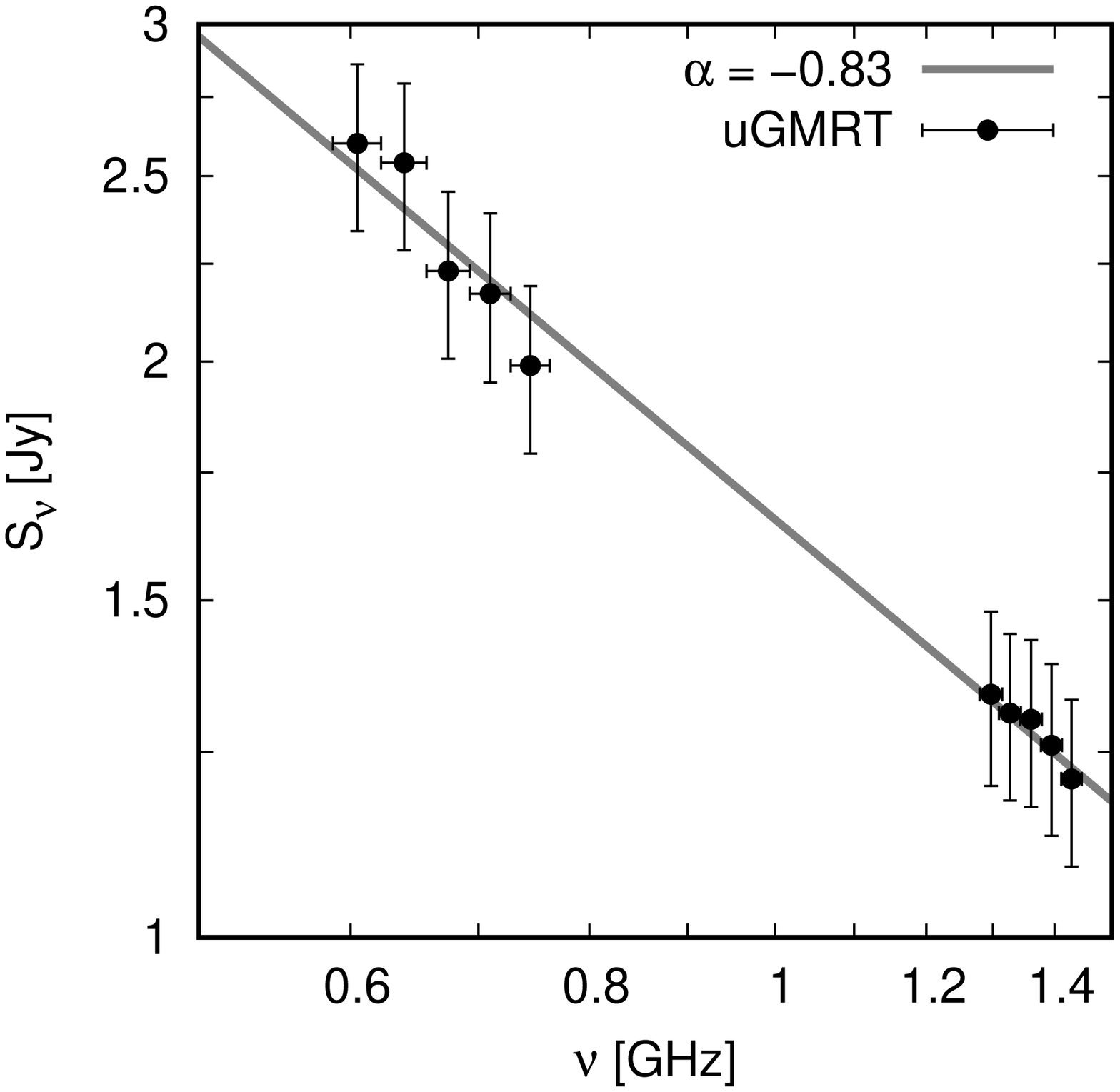}\\
    \vspace*{0.3cm}
    \includegraphics[angle=270, width=0.8\linewidth]{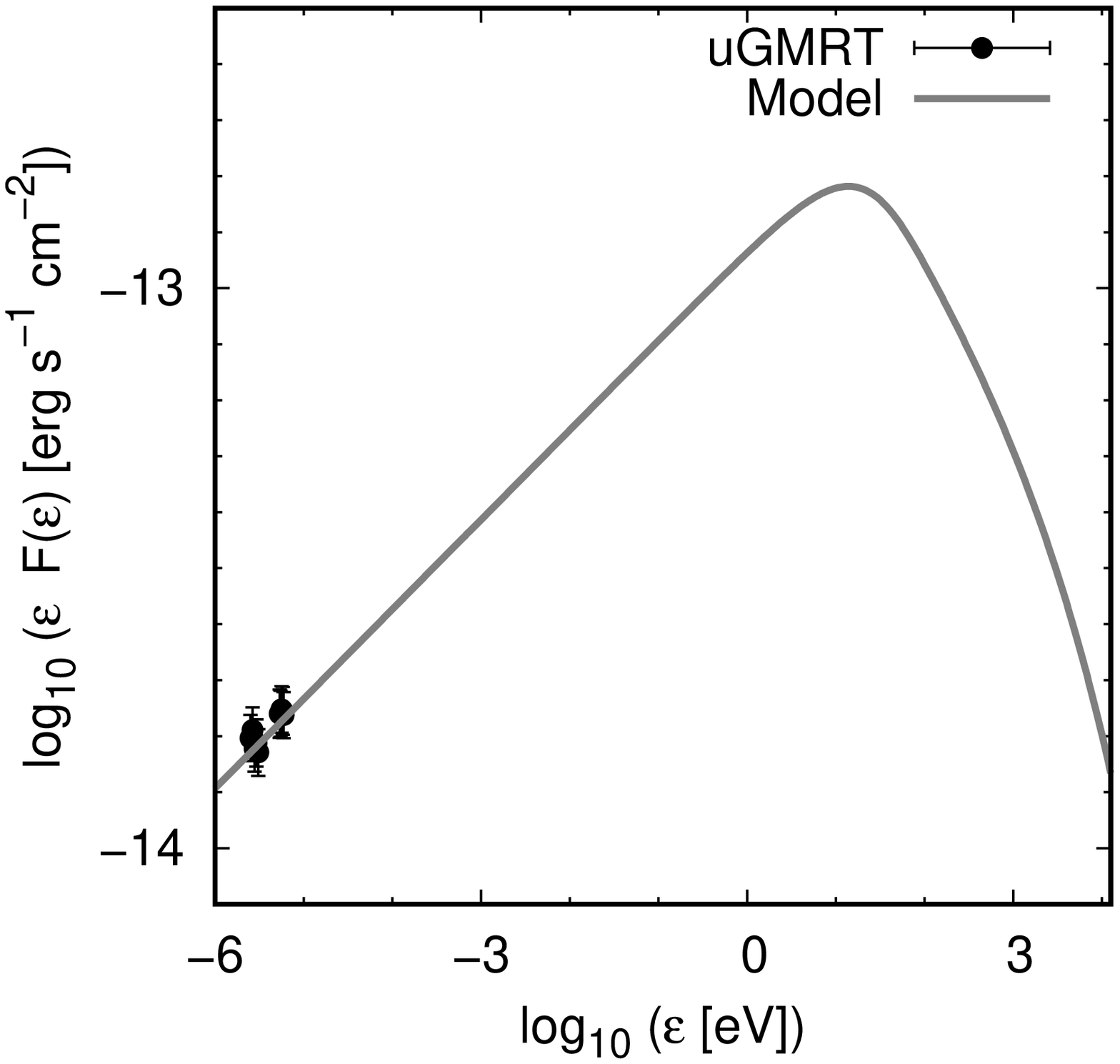}
%    \plottwo{./fit_alpha_log.eps}{./model_SED.eps}
    \caption{Top: Radio SED of G2.4+1.4 as obtained from ten uGMRT sub-bands. The best fit is shown which yields $\alpha = -0.83\pm0.1$. Bottom: Modeled synchrotron considering a one-zone emitter and a minimum energy condition for the magnetic field and the non-thermal particle population.}
    \label{fig:sed_model}
\end{figure}

Given that we have only two free parameters in our model, we explore physically plausible values for $\eta_\mathrm{mag}$ in the range 0.01 (strong magnetic field) to 10 (weak magnetic field) and obtain the corresponding value for $\chi_\mathrm{NT}$ that matches the observed fluxes. We calculate the synchrotron spectrum using the standard (approximated) expressions for electron emission in random magnetic fields \citep[e.g.][]{Melrose1980}. For the extreme case of ${\eta}_\mathrm{mag}=10$ we obtain $B = 120~\mu$G and $L_\mathrm{NT}=8\times {10}^{36}$~erg~s$^{-1}$ ($\chi_\mathrm{NT}\simeq 17\%$), whereas for $\eta_\mathrm{mag}=0.01$ we obtain $B=770~\mu$G and $L_\mathrm{NT}=0.3\times {10}^{36}$~erg~s$^{-1}$ ($\chi_{NT}\simeq 0.6\%$). A representative value of $\eta_\mathrm{mag}=0.75$, which corresponds to a minimum energy condition in the emitter, yields $B=250~\mu$G and $L_\mathrm{NT}=2\times {10}^{36}$~erg~s$^{-1}$ (${\chi}_{NT}\simeq 5\%$). In Figure~\ref{fig:sed_model} we show an example of the spectral fit for this case, which also allows to infer the possible X-ray flux due to the high end of the synchrotron spectrum. The expected X-ray flux in the 0.3--10~keV energy band is of the order of $10^{-13}$~erg~s$^{-1}$~cm$^{-2}$, though this value is very sensitive to uncertainties in the model such as spatial variations of the spectral index. Moreover, we note that thermal emission is also expected in the X-ray domain given the presence of shock-heated gas \citep{Toala2015,Toala2018}.

%==================================================================================================
%
\section{Conclusions}
%
%==================================================================================================

We report the first detection of non-thermal radiation from a single stellar bubble. This finding constitutes an observational breakthrough which gives new insight on the non-thermal physical processes taking place in the environments of isolated massive stars. In particular, this proves that non-runaway isolated massive stars are in fact capable of accelerating relativistic particles and are therefore confirmed as sources of Galactic cosmic rays. Nonetheless, their efficiency as cosmic ray accelerators is poorly constrained. We estimated the average magnetic field strength in the synchrotron emitting region using a model that relies in simple and general energy density partition considerations. Such a model suggests that the fraction of the available kinetic wind power that is converted into non-thermal particle acceleration is of the order of a few per cent. Future works devoted to a detailed modelling of the acceleration of cosmic rays in the termination shocks of stellar winds and their broadband emission will allow to place tighter constraints to this value.

%=========================================================
\section[]{Acknowledgments}
We thank the referee for useful suggestions.
We thank the staff of the GMRT, who made these observations possible. GMRT is run by the
National Centre for Radio Astrophysics of the Tata Institute of Fundamental Research. S.d.P. acknowledges financial support from CONICET (PIP 2014-00338) and thanks V. Bosch-Ramon for fruitful discussions. 
\bibliographystyle{aasjournal}
\bibliography{reference}

\begin{thebibliography}{}
\expandafter\ifx\csname natexlab\endcsname\relax\def\natexlab#1{#1}\fi
\providecommand{\url}[1]{\href{#1}{#1}}
\providecommand{\dodoi}[1]{doi:~\href{http://doi.org/#1}{\nolinkurl{#1}}}
\providecommand{\doeprint}[1]{\href{http://ascl.net/#1}{\nolinkurl{http://ascl.net/#1}}}
\providecommand{\doarXiv}[1]{\href{https://arxiv.org/abs/#1}{\nolinkurl{https://arxiv.org/abs/#1}}}

\bibitem[{{Arnal}(2008)}]{2008RMxAC..33..140A}
{Arnal}, E.~M. 2008, in Revista Mexicana de Astronomia y Astrofisica Conference
  Series, Vol.~33, Revista Mexicana de Astronomia y Astrofisica Conference
  Series, 140--141

\bibitem[{{Benaglia} {et~al.}(2010){Benaglia}, {Romero}, {Mart{\'{\i}}},
  {Peri}, \& {Araudo}}]{2010A&A...517L..10B}
{Benaglia}, P., {Romero}, G.~E., {Mart{\'{\i}}}, J., {Peri}, C.~S., \&
  {Araudo}, A.~T. 2010, \aap, 517, L10, \dodoi{10.1051/0004-6361/201015232}

\bibitem[{{Brighenti} \& {D'Ercole}(1995{\natexlab{a}})}]{1995IAUS..163...70B}
{Brighenti}, F., \& {D'Ercole}, A. 1995{\natexlab{a}}, in IAU Symposium, Vol.
  163, Wolf-Rayet Stars: Binaries; Colliding Winds; Evolution, ed. K.~A. {van
  der Hucht} \& P.~M. {Williams}, 70

\bibitem[{{Brighenti} \& {D'Ercole}(1995{\natexlab{b}})}]{1995MNRAS.273..443B}
{Brighenti}, F., \& {D'Ercole}, A. 1995{\natexlab{b}}, \mnras, 273, 443,
  \dodoi{10.1093/mnras/273.2.443}

\bibitem[{{De Becker}(2007)}]{DeBecker2007}
{De Becker}, M. 2007, \aapr, 14, 171, \dodoi{10.1007/s00159-007-0005-2}

\bibitem[{{De Becker}(2018)}]{DeBecker2018}
---. 2018, \aap, 620, A144, \dodoi{10.1051/0004-6361/201834101}

\bibitem[{{De Becker} \& {Raucq}(2013)}]{2013A&A...558A..28D}
{De Becker}, M., \& {Raucq}, F. 2013, \aap, 558, A28,
  \dodoi{10.1051/0004-6361/201322074}

\bibitem[{{De Becker} {et~al.}(2006){De Becker}, {Rauw}, {Sana}, {Pollock},
  {Pittard}, {Blomme}, {Stevens}, \& {van Loo}}]{2006MNRAS.371.1280D}
{De Becker}, M., {Rauw}, G., {Sana}, H., {et~al.} 2006, \mnras, 371, 1280,
  \dodoi{10.1111/j.1365-2966.2006.10746.x}

\bibitem[{{del Palacio} {et~al.}(2018){del Palacio}, {Bosch-Ramon},
  {M{\"u}ller}, \& {Romero}}]{delPalacio2018}
{del Palacio}, S., {Bosch-Ramon}, V., {M{\"u}ller}, A.~L., \& {Romero}, G.~E.
  2018, A\&A, 617, A13, \dodoi{10.1051/0004-6361/201833321}

\bibitem[{{del Valle} \& {Romero}(2012)}]{delValle2012}
{del Valle}, M.~V., \& {Romero}, G.~E. 2012, A\&A, 543, A56,
  \dodoi{10.1051/0004-6361/201218937}

\bibitem[{{del Valle} \& {Romero}(2014)}]{delValle2014}
---. 2014, A\&A, 563, A96, \dodoi{10.1051/0004-6361/201322308}

\bibitem[{{Dopita} \& {Lozinskaia}(1990)}]{1990ApJ...359..419D}
{Dopita}, M.~A., \& {Lozinskaia}, T.~A. 1990, \apj, 359, 419,
  \dodoi{10.1086/169074}

\bibitem[{{Dopita} {et~al.}(1990){Dopita}, {Lozinskaia}, {McGregor}, \&
  {Rawlings}}]{1990ApJ...351..563D}
{Dopita}, M.~A., {Lozinskaia}, T.~A., {McGregor}, P.~J., \& {Rawlings}, S.~J.
  1990, \apj, 351, 563, \dodoi{10.1086/168493}

\bibitem[{{Dorfi} \& {Drury}(1985)}]{1985ICRC....3..136D}
{Dorfi}, E.~A., \& {Drury}, L.~O. 1985, International Cosmic Ray Conference, 3

\bibitem[{{Dougherty} \& {Williams}(2000)}]{2000MNRAS.319.1005D}
{Dougherty}, S.~M., \& {Williams}, P.~M. 2000, \mnras, 319, 1005,
  \dodoi{10.1046/j.1365-8711.2000.03837.x}

\bibitem[{{Drury}(1983)}]{Drury1983}
{Drury}, L.~O. 1983, Reports on Progress in Physics, 46, 973,
  \dodoi{10.1088/0034-4885/46/8/002}

\bibitem[{{Dubner} \& {Giacani}(2015)}]{2015A&ARv..23....3D}
{Dubner}, G., \& {Giacani}, E. 2015, \aapr, 23, 3,
  \dodoi{10.1007/s00159-015-0083-5}

\bibitem[{{Eichler} \& {Usov}(1993)}]{Eichler1993}
{Eichler}, D., \& {Usov}, V. 1993, \apj, 402, 271, \dodoi{10.1086/172130}

\bibitem[{{Goss} \& {Lozinskaya}(1995)}]{1995ApJ...439..637G}
{Goss}, W.~M., \& {Lozinskaya}, T.~A. 1995, \apj, 439, 637,
  \dodoi{10.1086/175203}

\bibitem[{{Goss} \& {Shaver}(1968)}]{1968ApJ...154L..75G}
{Goss}, W.~M., \& {Shaver}, P.~A. 1968, \apjl, 154, L75, \dodoi{10.1086/180273}

\bibitem[{{Gray}(1994)}]{1994MNRAS.270..835G}
{Gray}, A.~D. 1994, \mnras, 270, 835, \dodoi{10.1093/mnras/270.4.835}

\bibitem[{{Green} \& {Downes}(1987)}]{1987MNRAS.225..221G}
{Green}, D.~A., \& {Downes}, A.~J.~B. 1987, \mnras, 225, 221,
  \dodoi{10.1093/mnras/225.2.221}

\bibitem[{Gupta {et~al.}(2017)Gupta, Ajithkumar, Kale, Nayak, Sabhapathy,
  Sureshkumar, Swami, Chengalur, Ghosh, Ishwara-Chandra, Joshi, Kanekar, Lal,
  \& Roy}]{article}
Gupta, Y., Ajithkumar, B., Kale, H., {et~al.} 2017, Current Science, 113, 707,
  \dodoi{10.18520/cs/v113/i04/707-714}

\bibitem[{{Haslam} {et~al.}(1982){Haslam}, {Salter}, {Stoffel}, \&
  {Wilson}}]{1982A&AS...47....1H}
{Haslam}, C.~G.~T., {Salter}, C.~J., {Stoffel}, H., \& {Wilson}, W.~E. 1982,
  \aaps, 47, 1

\bibitem[{{Johnson}(1975)}]{1975ApJ...198..111J}
{Johnson}, H.~M. 1975, \apj, 198, 111, \dodoi{10.1086/153581}

\bibitem[{Lal \& Rao(2006)}]{10.1111/j.1365-2966.2006.11225.x}
Lal, D.~V., \& Rao, A.~P. 2006, Monthly Notices of the Royal Astronomical
  Society, 374, 1085, \dodoi{10.1111/j.1365-2966.2006.11225.x}

\bibitem[{{Longair}(1992)}]{1992hea..book.....L}
{Longair}, M.~S. 1992, {High energy astrophysics. Vol.1: Particles, photons and
  their detection}, 436

\bibitem[{{Melrose}(1980)}]{Melrose1980}
{Melrose}, D.~B. 1980, {Plasma astrohysics. Nonthermal processes in diffuse
  magnetized plasmas - Vol.1: The emission, absorption and transfer of waves in
  plasmas; Vol.2: Astrophysical applications}

\bibitem[{Merten {et~al.}(2017)Merten, Becker~Tjus, Eichmann, \&
  Dettmar}]{Merten:2017mzg}
Merten, L., Becker~Tjus, J., Eichmann, B., \& Dettmar, R.-J. 2017, Astropart.
  Phys., 90, 75, \dodoi{10.1016/j.astropartphys.2017.02.007}

\bibitem[{{Mirabel} \& {Rodr{\'{\i}}guez}(1994)}]{Mirabel1994}
{Mirabel}, I.~F., \& {Rodr{\'{\i}}guez}, L.~F. 1994, Nature, 371, 46,
  \dodoi{10.1038/371046a0}

\bibitem[{{Polcaro} {et~al.}(1992){Polcaro}, {Viotti}, {Rossi}, \&
  {Norci}}]{1992A&A...265..563P}
{Polcaro}, V.~F., {Viotti}, R., {Rossi}, C., \& {Norci}, L. 1992, \aap, 265,
  563

\bibitem[{{Reich} {et~al.}(1984){Reich}, {Fuerst}, {Haslam}, {Steffen}, \&
  {Reif}}]{1984A&AS...58..197R}
{Reich}, W., {Fuerst}, E., {Haslam}, C.~G.~T., {Steffen}, P., \& {Reif}, K.
  1984, \aaps, 58, 197

\bibitem[{{Reich} {et~al.}(1990){Reich}, {Fuerst}, {Reich}, \&
  {Reif}}]{1990A&AS...85..633R}
{Reich}, W., {Fuerst}, E., {Reich}, P., \& {Reif}, K. 1990, \aaps, 85, 633

\bibitem[{{Reynolds}(2011)}]{2011Ap&SS.336..257R}
{Reynolds}, S.~P. 2011, \apss, 336, 257, \dodoi{10.1007/s10509-010-0559-8}

\bibitem[{{Rodr{\'{\i}}guez-Kamenetzky}
  {et~al.}(2019){Rodr{\'{\i}}guez-Kamenetzky}, {Carrasco-Gonz{\'a}lez},
  {Gonz{\'a}lez-Mart{\'{\i}}n}, {Araudo}, {Rodr{\'{\i}}guez}, {Vig}, \&
  {Hofner}}]{2019MNRAS.482.4687R}
{Rodr{\'{\i}}guez-Kamenetzky}, A., {Carrasco-Gonz{\'a}lez}, C.,
  {Gonz{\'a}lez-Mart{\'{\i}}n}, O., {et~al.} 2019, \mnras, 482, 4687,
  \dodoi{10.1093/mnras/sty3055}

\bibitem[{{Roger} {et~al.}(1999){Roger}, {Costain}, {Landecker}, \&
  {Swerdlyk}}]{1999A&AS..137....7R}
{Roger}, R.~S., {Costain}, C.~H., {Landecker}, T.~L., \& {Swerdlyk}, C.~M.
  1999, \aaps, 137, 7, \dodoi{10.1051/aas:1999239}

\bibitem[{{Rybicki} \& {Lightman}(1979)}]{1979rpa..book.....R}
{Rybicki}, G.~B., \& {Lightman}, A.~P. 1979, {Radiative processes in
  astrophysics}

\bibitem[{{S{\'a}nchez-Monge} {et~al.}(2013){S{\'a}nchez-Monge}, {Kurtz},
  {Palau}, {Estalella}, {Shepherd}, {Lizano}, {Franco}, \&
  {Garay}}]{Sanchez-Monge2013}
{S{\'a}nchez-Monge}, {\'A}., {Kurtz}, S., {Palau}, A., {et~al.} 2013, \apj,
  766, 114, \dodoi{10.1088/0004-637X/766/2/114}

\bibitem[{{Sander} {et~al.}(2018){Sander}, {Hamann}, {Todt}, {Hainich},
  {Shenar}, {Ramachandran}, \& {Oskinova}}]{2018arXiv180704293S}
{Sander}, A.~A.~C., {Hamann}, W.-R., {Todt}, H., {et~al.} 2018, ArXiv e-prints.
\newblock \doarXiv{1807.04293}

\bibitem[{{Seo} {et~al.}(2018){Seo}, {Kang}, \& {Ryu}}]{2018JKAS...51...37S}
{Seo}, J., {Kang}, H., \& {Ryu}, D. 2018, Journal of Korean Astronomical
  Society, 51, 37, \dodoi{10.5303/JKAS.2018.51.2.37}

\bibitem[{{Swarup} {et~al.}(1991){Swarup}, {Ananthakrishnan}, {Kapahi}, {Rao},
  {Subrahmanya}, \& {Kulkarni}}]{1991CuSc...60...95S}
{Swarup}, G., {Ananthakrishnan}, S., {Kapahi}, V.~K., {et~al.} 1991, Current
  Science, Vol.~60, NO.2/JAN25, P.~95, 1991, 60, 95

\bibitem[{{Toal{\'a}} \& {Arthur}(2018)}]{Toala2018}
{Toal{\'a}}, J.~A., \& {Arthur}, S.~J. 2018, \mnras, 478, 1218,
  \dodoi{10.1093/mnras/sty1127}

\bibitem[{{Toal{\'a}} {et~al.}(2015{\natexlab{a}}){Toal{\'a}}, {Guerrero},
  {Chu}, {Arthur}, \& {Gruendl}}]{Toala2015}
{Toal{\'a}}, J.~A., {Guerrero}, M.~A., {Chu}, Y.~H., {Arthur}, S.~J., \&
  {Gruendl}, R.~A. 2015{\natexlab{a}}, in Wolf-Rayet Stars: Proceedings of an
  International Workshop held in Potsdam, 333--336.
\newblock \doarXiv{1511.00861}

\bibitem[{{Toal{\'a}} {et~al.}(2015{\natexlab{b}}){Toal{\'a}}, {Guerrero},
  {Ramos-Larios}, \& {Guzm{\'a}n}}]{2015A&A...578A..66T}
{Toal{\'a}}, J.~A., {Guerrero}, M.~A., {Ramos-Larios}, G., \& {Guzm{\'a}n}, V.
  2015{\natexlab{b}}, \aap, 578, A66, \dodoi{10.1051/0004-6361/201525706}

\bibitem[{{Tramper} {et~al.}(2015){Tramper}, {Straal}, {Gr{\"a}fener}, {Kaper},
  {de Koter}, {Langer}, {Sana}, \& {Vink}}]{2015IAUS..307..144T}
{Tramper}, F., {Straal}, S.~M., {Gr{\"a}fener}, G., {et~al.} 2015, in IAU
  Symposium, Vol. 307, New Windows on Massive Stars, ed. G.~{Meynet},
  C.~{Georgy}, J.~{Groh}, \& P.~{Stee}, 144--145,
  \dodoi{10.1017/S1743921314006590}

\bibitem[{{Treffers} \& {Chu}(1982)}]{1982ApJ...254..132T}
{Treffers}, R.~R., \& {Chu}, Y.-H. 1982, \apj, 254, 132, \dodoi{10.1086/159715}

\bibitem[{{Vig} {et~al.}(2018){Vig}, {Veena}, {Mandal}, {Tej}, \&
  {Ghosh}}]{2018MNRAS.474.3808V}
{Vig}, S., {Veena}, V.~S., {Mandal}, S., {Tej}, A., \& {Ghosh}, S.~K. 2018,
  \mnras, 474, 3808, \dodoi{10.1093/mnras/stx3032}

\bibitem[{{Wright} \& {Barlow}(1975)}]{1975MNRAS.170...41W}
{Wright}, A.~E., \& {Barlow}, M.~J. 1975, \mnras, 170, 41,
  \dodoi{10.1093/mnras/170.1.41}

\bibitem[{{Yang} {et~al.}(2018){Yang}, {de O{\~n}a Wilhelmi}, \&
  {Aharonian}}]{2018A&A...611A..77Y}
{Yang}, R.-z., {de O{\~n}a Wilhelmi}, E., \& {Aharonian}, F. 2018, \aap, 611,
  A77, \dodoi{10.1051/0004-6361/201732045}

\end{thebibliography}
\end{document}